\documentclass[aps,prc,twocolumn,superscriptaddress,floatfix,showpacs,nofootinbib]{revtex4-2}
\usepackage{multirow}
\usepackage{graphicx}
\usepackage{color}
\usepackage{amsmath}
\usepackage{amssymb}
\usepackage[colorlinks=true]{hyperref}
\usepackage{tikz}
\usepackage{etoolbox}
\usepackage{bm}
\usepackage[utf8]{inputenc}
\usepackage{cleveref}
\usepackage{amsmath}
\usepackage{amssymb}
\usepackage{verbatim}
\usepackage{float}
\usepackage[bottom]{footmisc}

\newcommand {\snn}      {\sqrt{s_{_{\rm NN}}}}

\newcommand {\trento}   {{\tt TRENTo}}

\newcommand {\urqmd}    {{\tt UrQMD}}
\newcommand {\iebe}   {{\tt iEBE-VISHNU}}

\newcommand {\vis}      {{\tt VISH2+1}}

\newcommand{\n}{\nonumber}

\newcommand {\ox}   {$^{16}$O}
\newcommand {\oo}   {$^{16}$O+$^{16}$O}

\begin{document}

\title{Exploring the compactness of $\alpha$ cluster in $^{16}$O nuclei with relativistic $^{16}$O+$^{16}$O  collisions}

\author{Yuanyuan Wang}
\affiliation{School of Physics, Peking University, Beijing 100871, China}

\author{Shujun Zhao}
\affiliation{School of Physics, Peking University, Beijing 100871, China}

\author{Boxing Cao}
\affiliation{School of Physics, Peking University, Beijing 100871, China}

\author{Hao-jie Xu}
\email{haojiexu@zjhu.edu.cn}
\affiliation{School of Science, Huzhou University, Huzhou, Zhejiang 313000, China}
\affiliation{Strong-Coupling Physics International Research Laboratory (SPiRL), Huzhou University, Huzhou, Zhejiang 313000, China.}

\author{Huichao Song}
\email{huichaosong@pku.edu.cn}
\affiliation{School of Physics, Peking University, Beijing 100871, China}
\affiliation{Collaborative Innovation Center of Quantum Matter, Beijing 100871, China}
\affiliation{Center for High Energy Physics, Peking University, Beijing 100871, China}
\date{\today}
\begin{abstract}
Probing the $\alpha$ cluster of $^{16}$O with the relativistic \oo\ collisions has raised great interest in the heavy ion community.
However, the effects of the $\alpha$ cluster on the soft hadron observables vary largely among different studies.
In this paper, we explain the differences by the compactness of the $\alpha$ cluster in oxygen, using \iebe\ hydrodynamic simulations with different initial state $\alpha$ cluster configurations.
We also find several observables, such as the intensive skewness of the $[p_{\rm T}]$ correlator $\Gamma_{p_{\rm T}}$, the harmonic flows $v_2\{2\}$, $v_2\{4\}$, $v_3\{2\}$, and the $v_n^2-\delta[p_{\rm T}]$ correlations $\rho(v_{2}^{2}, [p_{\rm T}])$, $\rho(v_{3}^{2}, [p_{\rm T}])$ in \oo\ collisions are sensitive to the compactness of the $\alpha$ cluster in the colliding nuclei, which can be used to constrain the configurations of $^{16}$O in the future.
Our study serves as an important step toward the quantitative exploration of the $\alpha$ cluster configuration in the light nuclei with relativistic heavy ion collisions.
\end{abstract}

\maketitle

{\em Introduction.}
The configurations of $\alpha$ clusters in nuclei have attracted much attention from researchers for almost a hundred years since the idea was first proposed by Gamow~\cite{gamow1931constitution}. In particular, the triangular configurations in $^{12}$C and the tetrahedral configurations in \ox\ have been extensively discussed for decades~\cite{Hoyle:1954zz,Cook:1957zz,Dunbar:1953zz,ikeda1968systematic, Epelbaum:2013paa,Lahde:2013uqa, Bijker:2014tka, Bijker:2016bpb,Kanada-Enyo:2006rjf, DellAquila:2017ppe, Smith:2017jub}. 
Various approaches have been proposed to study these configurations in both the ground and excited states of the nucleus~\cite{Wheeler:1937zza,wildermuth1958cluster,Tohsaki:2001an,Feldmeier:1989st,Ono:1992uy,Navratil:2000gs,coester1958bound,Kummel:1978zz}. 
One of the most interesting approaches is the relativistic heavy ion collisions, where the structure information of the colliding nuclei is imprinted in the created quark-gluon plasma (QGP)~\cite{Broniowski:2013dia,Zhang:2017xda, Rybczynski:2017nrx, Guo:2019sek, Lim:2018huo, Rybczynski:2019adt,Wang:2021ghq, Li:2020vrg, Behera:2021zhi, Ding:2023ibq, Summerfield:2021oex,Huang:2023viw,Schenke:2020mbo,Nijs:2021clz,Liu:2023gun}. The original idea was to collide a light nucleus against a heavy nucleus at high energies to constrain the $\alpha$ cluster configurations of the light nucleus~\cite{Broniowski:2013dia}. Recently, both RHIC and the LHC have performed or have decided to perform \oo\ collisions at high energies~\cite{Huang:2023viw,Brewer:2021kiv}, providing opportunities to probe the $\alpha$ cluster configurations in a heavy ion experiment. 

In relativistic heavy ion collisions, the nuclei pass each other in a very short time. The spatial distribution of the colliding nuclei is recorded instantaneously in the initial stage of QGP, which leaves messages in the final state correlations of the emitted hadrons. Since the dynamic evolution of the QGP medium can be well described by relativistic hydrodynamics or transport approaches, the final state observables could be used to study the size and shape of the initial state. The best example of which is the relativistic isobaric collisions that the system uncertainties from the detectors and the bulk properties of the QGP medium can be largely canceled~\cite{Li:2019kkh,Xu:2021vpn,Xu:2021uar,Zhao:2022uhl}. The correlations of the initial nucleons in the $\alpha$ cluster nuclei can also be probed by the correlations of the final particles in the heavy ion collisions. The STAR preliminary results already give some insight into the configuration of \ox\ in relativistic \oo\ collisions~\cite{Huang:2023viw}.

Recently, tremendous efforts have been made to investigate the effect of $\alpha$ cluster on the flow harmonics and other observables in $^{16}$O+$^{16}$O collisions with the initial geometry models, the hydrodynamic models and the transport models~\cite{Lim:2018huo, Rybczynski:2019adt, Wang:2021ghq, Li:2020vrg, Behera:2021zhi, Ding:2023ibq, Summerfield:2021oex,Huang:2023viw,Schenke:2020mbo,Nijs:2021clz}. 
Most of these studies focus on a typical configuration of $\alpha$ clusters in \ox, obtained by nuclear structure theories or simple geometric constructions.
However, due to different Hamiltonians/arppoximations, calculations such as Variational Monte Carlo (VMC), Nuclear Lattice Effective Field Theory (NLEFT), and Extended Quantum Molecular Dynamics (EQMD) give very different tetrahedral-like clustering correlations~\cite{Epelbaum:2013paa,Rybczynski:2017nrx,Pieper:2002ne,Lee:2008fa,He:2014iqa}. These different structures lead to different predictions on the observables of relativistic \oo\ collisions~\cite{Huang:2023viw}. We find that these different predictions may be due to the compactness of the $\alpha$ cluster in \ox\ -- there could be a loose or compact "$\alpha$" in the nuclei, compared to the size of free $\alpha$ nuclei ($r_{\alpha}\equiv\sqrt{\langle r^2\rangle}= 1.71$ fm). In this work, we will use a state-of-the-art relativistic hydrodynamic model to study the effect of tetrahedral $\alpha$ cluster configurations on the final observables of \oo\ collisions at $\snn=6.5$ TeV.

{\em Model and setups.}
The dynamic evolution of the QGP medium created by the \oo\ collisions at $\snn=6.5$ TeV is simulated by an \iebe~\cite{Shen:2014vra,Song:2010aq} model. 
The \iebe\ is an event-by-event hybrid model that combines the \trento\ model generating the initial stage~\cite{Moreland:2014oya,Bernhard:2016tnd}, the \vis{} describing the collective expansion of the QGP~\cite{Heinz:2005bw,Song:2007ux,Song:2007fn}, and the \urqmd~\cite{Bass:1998ca, Bleicher:1999xi} simulating the evolution of the hadron cascade in the hadronic rescattering process. 
All the model parameters are listed in Tab.~\ref{tab:VISHNU}, except those for the structure of $^{16}$O. With these parameters, the anisotropic flow observables measured by the ALICE and CMS collaborations~\cite{ALICE:2015juo, CMS:2013jlh, ALICE:2013rdo} can be well described with our \iebe\ simulations.

\begin{figure}
\begin{center}
\includegraphics[width=0.45\textwidth]{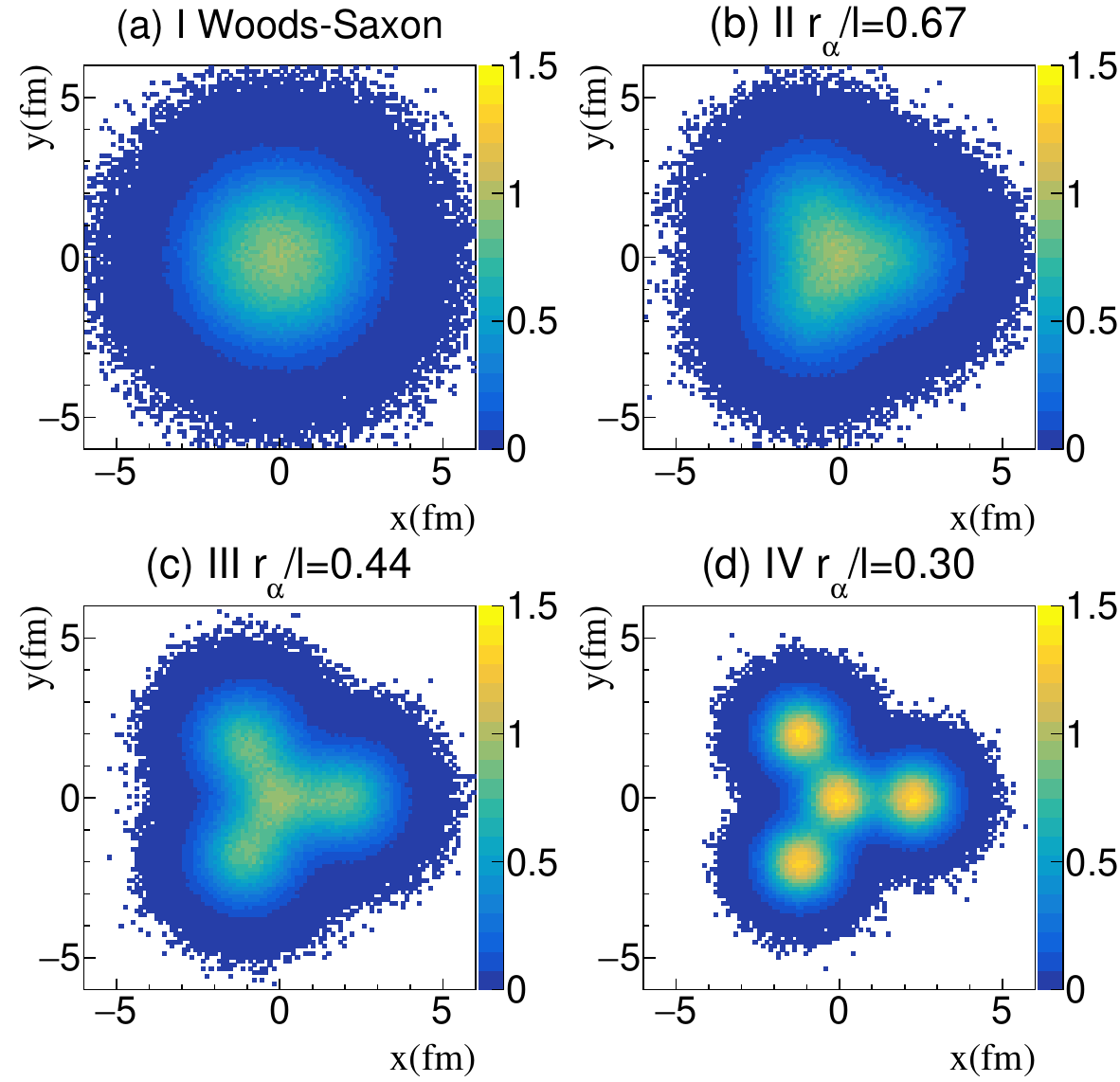}
	\caption{(Color online) The 2-dimensional density distributions (unit in fm$^{-2}$) of the different tetrahedral configurations of $^{16}$O listed in Tab.~\ref{tab:stru}, obtained by integrating their nuclear densities along one of their C3 rotation axes.}
\label{f:str}
\end{center}
\end{figure}

In this work, we only focus  on the tetrahedral configurations of $\alpha$ clusters in $^{16}$O, but with more detailed discussions on the compactness of the $\alpha$ in the nuclei. Some other configurations such as linear chain and Y-shape configurations are also of interest but are beyond the scope of this study. In the tetrahedral configurations, the shape of oxygen is described by a tetrahedron of side length $l$, and the centers of four $\alpha$ clusters are placed at the vertices of the tetrahedron. The spatial coordinates of the nucleons in each $\alpha$ cluster are sampled from a 3D Gaussian distribution with root-mean-square radius  $r_\alpha$, which describes the mean radius of each cluster. The $r_\alpha$ parameter reflects the compactness of the $\alpha$ cluster in the nuclei. A smaller $r_\alpha$ indicates a denser cluster in the nuclei. In this study, the nuclear density of $^{16}$O is constructed with three magnitudes of $r_\alpha$ to comprehensively understand the effect of $\alpha$ cluster configurations on the final observables. We enforce that the root-mean-square radius of $^{16}$O should be the same for the different densities ($\sqrt{\langle r^2\rangle}\equiv \sqrt{3l^2/8 + r_{\alpha}^2} = 2.73$ fm from the nuclear structure experiment~\cite{de1987c}, here we ignore the differences between charge density and nuclear mass density), a large $l$ is being required for the case of compacted $\alpha$ cluster. For comparison, a three parameter Fermi distribution (3pF) with the same root-mean-square radius of $^{16}$O is also computed, $\rho=\rho_0\left(1+\omega r^2/R^2\right)\left[1+\exp\left((r-R)/a\right)\right]^{-1}$, where $R=2. 608$ fm, $a=0.513$ fm, $\omega=-0.051$ fm~\cite{de1987c}. The corresponding parameters are listed in Tab.~\ref{tab:stru} and the 2-dimensional densities obtained by integrating their nuclear densities along one of their C3 rotation axes are shown in Fig.~\ref{f:str}.
Obvious triangular hotspots appear at small $r_{\alpha}/l$.
The one-body density $\rho(r)$ was enhanced at $r\sim2$ fm with small $r_{\alpha}/l$, consistent with recent study~\cite{Giacalone:2024luz}. We find that the two-body correlation functions $C(r)$~\cite{Rybczynski:2019adt} imply attractive effect at low separations $r$ with small $r_{\alpha}/l$, this needs to be further investigated with more realistic nuclear structure theory calculations.

\begin{table}
      \caption{The \iebe{} parameters for the simulation of \oo\ collisions at $\snn=6.5$ TeV. A detailed description of these parameters can be found Ref.~\cite{Moreland:2018gsh, Summerfield:2021oex}.
      \label{tab:VISHNU}}
      \centering{}%
    \begin{tabular}
    {p{1.7cm}p{1.8cm}p{2cm}p{2cm}}
    \hline
              \multicolumn{2}{c}{Initial condition/Preeq.} & \multicolumn{2}{c}{QGP medium}  \\
    \hline
          ${\rm Norm}$           & $17$ GeV      & $(\eta/s)_{\rm min}$           & $0.11$   \\
          $p$                   & $0.0$          & $(\eta/s)_{\rm slope}$           & $1.6$ GeV$^{-1}$    \\
          $\sigma_{\rm flut}$   & $1.6$       & $(\eta/s)_{\rm crv}$            & $-0.29$    \\
          $r_{\rm cp}$          & $0.51$ fm     & $(\zeta/s)_{\rm max}$         & $0.032$   \\
          $n_{\rm c} $             & $1$          & $(\zeta/s)_{\rm width}$       & $0.024$ GeV   \\
          $w_{\rm c}$              & $0.51$ fm     & $(\zeta/s)_{T_{0}}$    & $0.175$ GeV   \\
          $d_{\rm min}$        & $0.4$ fm       & $T_{\rm switch}$                & $0.151$ GeV   \\
          $\tau_{\rm fs}$        & $0.37$ fm/$c$ &                                             &    \\
      \hline
      \end{tabular}
\end{table}

\begin{table}
      \caption{ The parameters (side length $l$ of tetrahedron, the rms radius $r_\alpha$ of each cluster) for the nuclear distributions of \ox\ with tetrahedral configurations of $\alpha$ clusters. The parameters for the Woods-Saxon distribution is $R=2.608$ fm, $a=0.513$ fm, and $\omega=-0.051$ fm~\cite{de1987c}.\label{tab:stru}}
      \centering{}%
    \begin{tabular}
    {p{1cm}p{2.5cm}p{1.3cm}p{1.3cm}p{1.3cm}}
    \hline
            & distribution &  $l$ &   $r_\alpha$   &   $r_{\alpha}/l$   \\  
    \hline
        I    & Woods--Saxon       &      &    &     \\   
        II  & $\alpha$ cluster      & 3.0  & 2.0  & 0.67  \\  
        III   & $\alpha$ cluster      & 3.6  & 1.6  & 0.44  \\ 
        IV   & $\alpha$ cluster      & 4.0  & 1.2  & 0.30  \\  
    \hline
      \end{tabular}
\end{table}

We simulate $\sim50$ k hydrodynamic events of \oo\ collisions at $\snn=6.5$ TeV for each nuclear density at the top  $50\%$ centrality, together with 2000 oversamplings of \urqmd\ afterburner for each hydrodynamic event. The centrality is determined by the charged particle multiplicity with $|\eta|<0.5$.
Based on these simulations, we find that several observables are sensitive to the configurations of \ox, such as the mean transverse momenta $\langle p_{\rm T}\rangle\equiv\langle[p_{\rm T}]\rangle$, the two-particle $[p_{\rm T}]$ correlator $\langle \Delta p_{{\rm T}_i} \Delta p_{{\rm T}_j}\rangle$, the intensive skewness of $[p_{\rm T}]$ correlator $\Gamma_{p_{\rm T}}$, the elliptic flows $v_2\{2\}$, $v_2\{4\}$ and their ratios, the triangular flow $v_3\{2\}$, as well as the $v_n^2-\delta[p_{\rm T}]$ correlations $\rho(v_{2}^2, [p_{\rm T}])$ and $\rho(v_{3}^2, [p_{\rm T}])$. Here $[p_{\rm T}]$ is the mean transverse momentum of a given event and $\langle...\rangle$ denotes the average over the ensemble of events. 

Before the discussion of the results, some definitions of these observables are given below. The intensive skewness is defined by~\cite{Giacalone:2020lbm}
\begin{align}
    \Gamma_{p_{\rm T}}&=\frac{\langle \Delta p_{{\rm T}_i} \Delta p_{{\rm T}_j}\Delta p_{{\rm T}_k}\rangle\langle p_{\rm T}\rangle}{{\langle \Delta p_{{\rm T}_i} \Delta p_{{\rm T}_j}\rangle}^2}.
\end{align}
Here $\langle \Delta p_{{\rm T}_i} \Delta p_{{\rm T}_j}\rangle$ and $\langle \Delta p_{{\rm T}_i} \Delta p_{{\rm T}_j}\Delta p_{{\rm T}_k}\rangle$ are the two- and three-particle correlators of $[p_{\rm T}]$, defined as follows
\begin{align}
\langle \Delta p_{{\rm T}_i} \Delta p_{{\rm T}_j}\rangle & =\left\langle \frac{{Q_1}^2-Q_2}{N_{\rm ch}(N_{\rm ch}-1)}  \right\rangle-{\left\langle \frac{Q_1}{N_{\rm ch}}\right\rangle}^2 , \\
\langle \Delta p_{{\rm T}_i} \Delta p_{{\rm T}_j}\Delta p_{{\rm T}_k}\rangle & =\left\langle \frac{{Q_1}^3+2Q_3-3Q_1Q_2}{(N_{\rm ch}-1)(N_{\rm ch}-2)}\right\rangle + 2{\left\langle \frac{Q_1}{N_{\rm ch}}  \right\rangle}^3 \n\\
&-3 {\left\langle \frac{Q_1}{N_{\rm ch}} \right\rangle}\left\langle \frac{{Q_1}^2-Q_2}{N_{\rm ch}(N_{\rm ch}-1)}\right\rangle.
\end{align}
with $Q_n=\sum_{i=1}^{N_{\rm ch}}p^n_{{\rm T}, i}$.

The Pearson correlation coefficient of $v_n^2-\delta[p_{\rm T}]$ correlation is defined by~\cite{Bozek:2016yoj}
\begin{align}
\rho(v_n^2, [p_{\rm T}])&=\frac{\langle v_n^2\delta[p_{\rm T}]\rangle}{\sqrt{\langle(\delta v_n^2)^2\rangle\langle(\delta[p_{\rm T}])^2\rangle}},
\end{align}
which is an observable that sensitive to the initial geometry and its fluctuations. The associated flow harmonics are calculated with the Q-cumulant method~\cite{Bilandzic:2010jr}.

{\em Results and Discussions.}
For our hydrodynamic simulations, the centrality cuts are slightly different with different $^{16}$O densities.
However, for the observables discussed in this study, the bias due to centrality cut differences are negligible when comparing the four cases. 
Quantitatively, most of the results discussed in this work can be described by the related initial predictors, i.e., $[p_{\rm T}]\propto d_{\perp}\equiv E/S$ and $v_{n}\propto \epsilon_{n}$~\cite{Teaney:2010vd,Giacalone:2020dln}, see APPENDIX. Here $E$ and $S$ are the initial total energy and entropy, and $\epsilon_{n}$ are the initial eccentricities.

\begin{figure*}
\begin{center}
\includegraphics[width=0.98\textwidth]{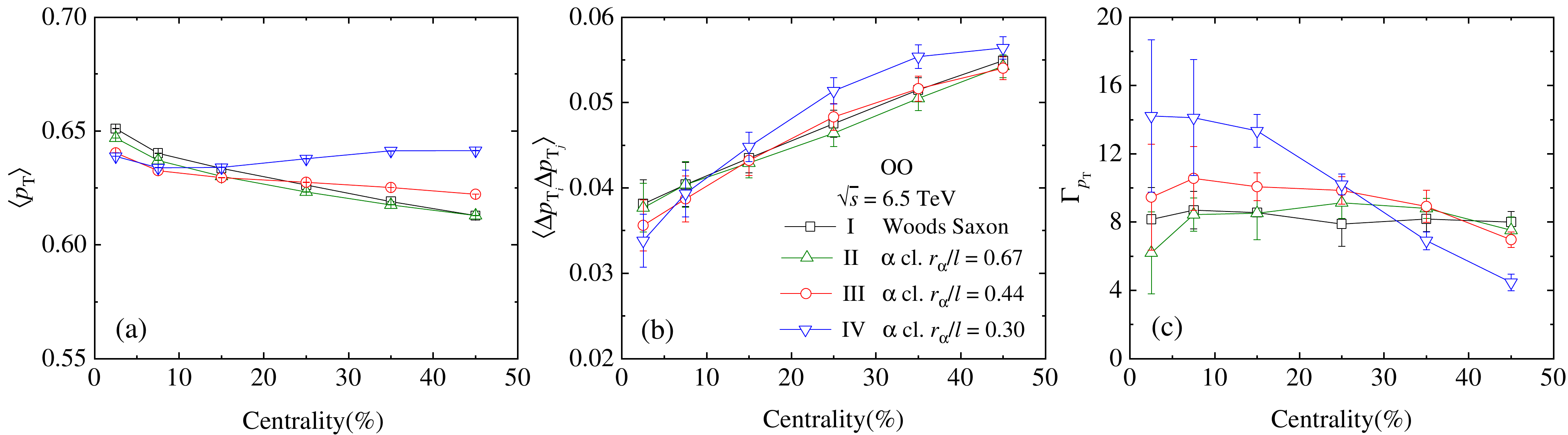}
	\caption{(Color online) The centrality dependent (a) mean transverse momenta $\langle p_{\rm T}\rangle$, (b) two-particle $[p_{\rm T}]$ correlator  $\langle \Delta p_{{\rm T}_i} \Delta p_{{\rm T}_j}\rangle $ and (c) intensive skewness of $[p_{\rm T}]$ correlator $\Gamma_{p_{\rm T}}$ of charged hadrons in \oo\ collisions at $\snn=6.5$ TeV, calculated by the \iebe\ model with different initial state $\alpha$ cluster configurations.}
\label{f:pt}
\end{center}
\end{figure*}

Fig.~\ref{f:pt} shows the centrality dependent cumulants of the $\langle p_{\rm T}\rangle$ distributions. Except for the very compact cluster in case IV, the predictions of these cumulants are roughly overlap within error bars. In relativistic heavy ion collisions, the $\langle p_{\rm T}\rangle$ depends on the density of the overlap region~\cite{Broniowski:2009fm,Schenke:2020uqq}, and the magnitude typically  decreases with centrality, as in cases I-III shown in Fig.~\ref{f:pt}(a). However, when the $\alpha$ cluster is highly compact in the nuclei, as in case IV, there is a non-monotonic centrality dependence for the $\langle p_{\rm T}\rangle$, as the compact cluster reduces the $\langle p_{\rm T}\rangle$ for central collisions and increases it for peripheral collisions. The effect of $\alpha$ configurations on two-particle $[p_{\rm T}]$ correlator $\langle \Delta p_{{\rm T}_i} \Delta p_{{\rm T}_j}\rangle $ shown in Fig.~\ref{f:pt}(b) is similar to the $\langle p_{\rm T}\rangle$, except that all trends show the correlator increasing with centrality. Conversely, the compact $\alpha$ cluster increases the intensity of the $[p_{\rm T}]$ correlator $\Gamma_{p_{\rm T}}$ at central collisions and decreases it at peripheral collisions, see Fig.~\ref{f:pt}(c). The centrality dependence of $\Gamma_{p_{\rm T}}$ is weak for the first three cases of nuclear densities, while its prediction from case IV nuclear density shows a very obvious centrality dependence, the value decreasing with centrality.

\begin{figure*}
\begin{center}
\includegraphics[width=0.98\textwidth]{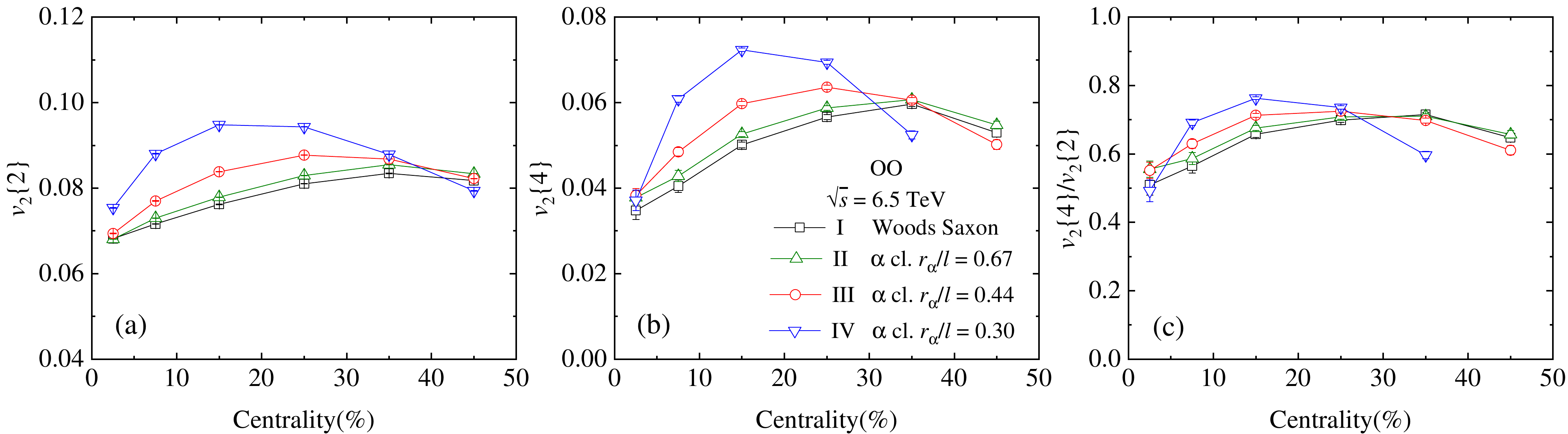}
	\caption{(Color online) The centrality dependent (a) $v_2\{2\}$, (b) $v_2\{4\}$, and (c) their ratios of all charged hadrons in $^{16}$O+$^{16}$O collisions at $\sqrt{s_{\rm{NN}}}=6.5$ TeV, calculated by  the \iebe\ model.}
\label{f:v2}
\end{center}
\end{figure*}

\begin{figure}
\begin{center}
\includegraphics[width=0.42\textwidth]{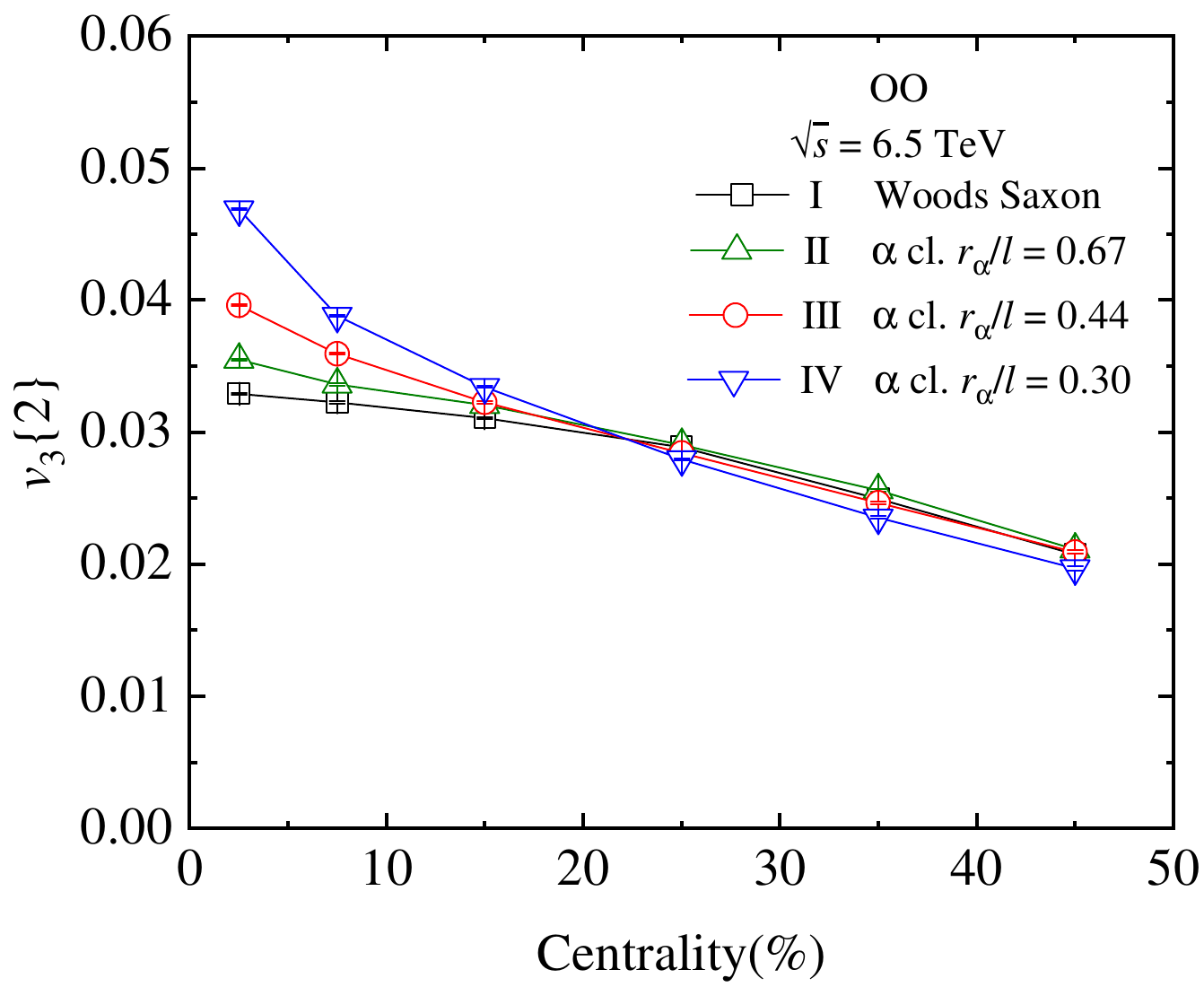}
\caption{(Color online) The centrality dependent $v_3\{2\}$ of all charged hadrons in $^{16}$O+$^{16}$O collisions at $\sqrt{s_{\rm{NN}}}=6.5$ TeV, calculated by the \iebe\ model.}
\label{f:v32}
\end{center}
\end{figure}

The effect of $\alpha$ cluster on the flow observables in \oo\ collisions are shown in Fig.~\ref{f:v2} and Fig.~\ref{f:v32}.
Such effect has been studied extensively in previous work. 
However, different studies give different conclusions on the $\alpha$ cluster effect. Some of them predicted that the effect of $\alpha$ cluster on flow observables is considerably small~\cite{Summerfield:2021oex,Huang:2023viw}, while some other studies indicate that the flow observables can be used to detect the $\alpha$ cluster in oxygen~\cite{Rybczynski:2019adt,Li:2020vrg,Huang:2023viw}. This may be due to the different configurations used in their models. Here, we give a possible way to understand these differences.

Fig.~\ref{f:v2} shows  the centrality dependent $v_{2}\{2\}$, $v_{2}\{4\}$ and their ratios in \oo\ collisions at $\snn=6.5$ TeV, calculated from \iebe\ with different $\alpha$ cluster configurations~\footnote{ $c_{2}\{4\}$ tunes to 
positive  value in the $40-50\%$ centrality, there thus  is no  $v_{2}\{4\}\equiv(-c_{2}\{4\})^{1/4}$ value above $40-50\%$ centrality for case IV in Fig.~\ref{f:v2}(b,c)}. The predictions from the Woods-Saxon density (case I) and the loose cluster density (case II) are similar. The reason is that with the large $r_\alpha/l$ in case II, the four $\alpha$ clusters in oxygen overlap with each other, and we get a smooth nuclear density as in the Woods-Saxon case.  
The compact $\alpha$ (case IV), however, introduces more fluctuations into the initial state, giving very different predictions for the centrality dependent elliptic flow. Especially for mid-center collisions like $10-30\%$ centralities, the enhancements of $v_{2}\{2\}$ and $v_{2}\{4\}$ due to the compact cluster in oxygen are obvious. An interesting feature is that such enhancements are larger for the elliptical flow obtained from four-particle correlations than those from two-particle correlations, resulting in non-trivial centrality dependent $v_{2}\{4\}/v_{2}\{2\}$ ratios, as shown in Fig.~\ref{f:v2}(c).

For flow observables in a single collision system, their individual magnitude depends on the properties of the QGP medium, we therefore prefer to discuss their ratios to explore the nuclear structure effect~\cite{Xu:2021uar,Liu:2023pav}. The $v_{2}\{4\}/v_{2}\{2\}$ have been used to study the $\alpha$ configurations in heavy ion experiments. In comparison to the initial model simulations, the STAR preliminary results on the centrality dependent $v_{2}\{4\}/v_{2}\{2\}$ ratio in central collisions are consistent with the prediction with $\alpha$ configuration from VMC calculations, while the prediction with $\alpha$ configuration from NLEFT calculations somehow failed. Based on our study with hydrodynamic simulations, the different predictions on the trends of the centrality dependent $v_{2}\{4\}/v_{2}\{2\}$ ratios are due to the different $r_\alpha/l$ under tetrahedral $\alpha$ configurations, the hydrodynamic simulations with a smaller $r_\alpha/l$ predict a rapid increase at the top $10\%$ centrality. From nuclear structure theories, we know that the effective $r_\alpha/l$ from VMC is smaller than the one from NLEFT~\cite{Epelbaum:2013paa,Rybczynski:2017nrx}, consistent with our conclusions. Our study of the top RHIC energy is ongoing. 

Fig.~\ref{f:v32} shows the centrality dependent triangular flow $v_{3}\{2\}$ in \oo\ collisions at $\snn=6.5$ TeV, calculated by \iebe\ model. The effect of the $\alpha$ cluster on $v_{3}\{2\}$ is obvious in the most central collisions. As mentioned above, the compact $\alpha$ cluster contributes large fluctuations to the initial profiles. However, one would expect such an effect to introduce some enhancement of the $v_{3}\{2\}$ for the whole centrality range. Therefore, the only enhancement at most central collisions shown in Fig.~\ref{f:v32} is most likely due to the geometry becoming dominant contributions.
We know that a large octupole deformation is an enhancement of $v_{2}\{2\}$ in mid-central collisions and $v_{3}\{2\}$ in most-central collisions~\cite{jia:2021tzt}, which is similar to the effect of the $\alpha$ clusters shown in Fig.~\ref{f:v2} and Fig.~\ref{f:v32}. In fact, the cluster structure in oxygen indicates finite $Q_{3X}$ with octupole deformation~\cite{Wang:2019dpl}. If we project the density into the transverse plane of the heavy ion collisions, we get an obvious triangular structure like the Hoyle $^{12}$C, as shown in Fig.~\ref{f:str}. Therefore, a non-zero $\beta_{3}$ would be required to parameterize the clustered oxygen with the Woods-Saxon formula, and it is interesting to further investigate the differences between the clustered density and its Woods-Saxon parameterization.

\begin{figure*}
\begin{center}
\includegraphics[width=0.9\textwidth]{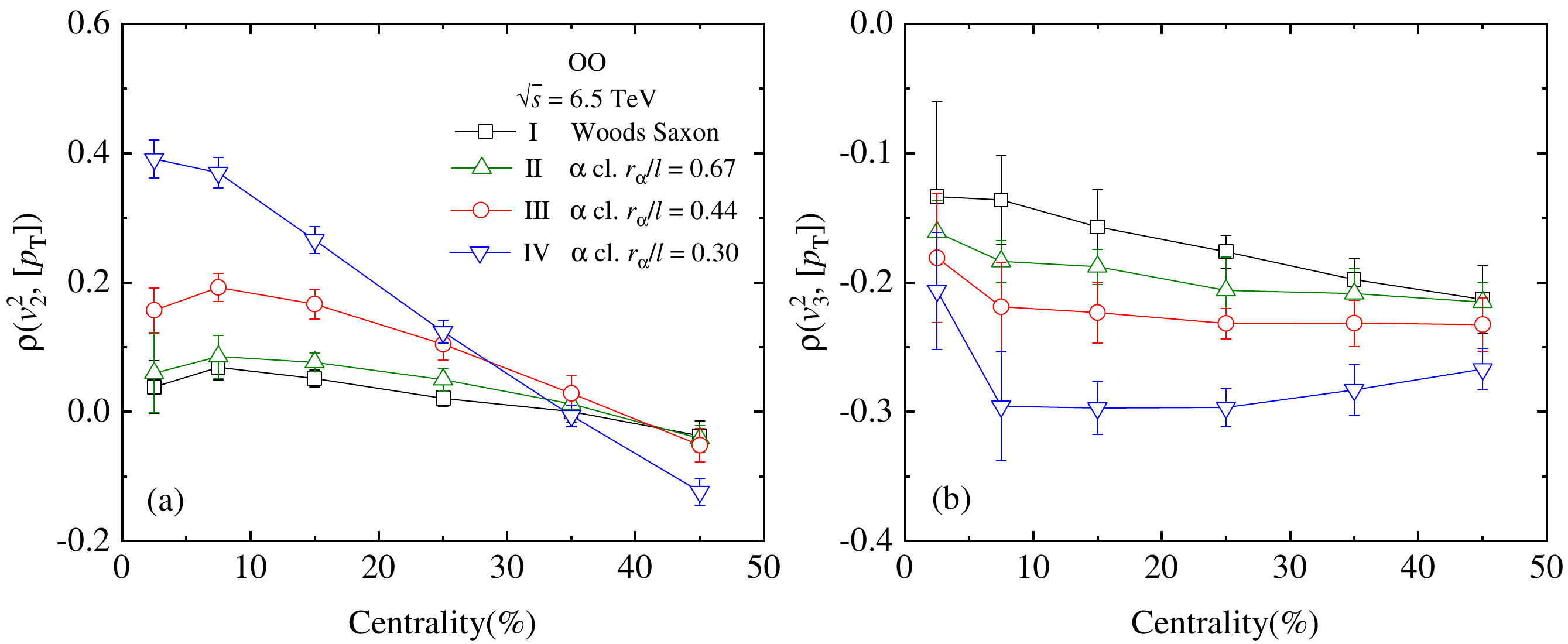}
	\caption{(Color online) The charged hadron Pearson correlation coefficients (a) $\rho(v_{2}^{2}, [p_{\rm T}])$ and (b) $\rho(v_{3}^{2}, [p_{\rm T}])$ as a function of centrality in $^{16}$O+$^{16}$O collisions at $\sqrt{s_{\rm{NN}}}=6.5$ TeV, calculated by the \iebe\ model.}
\label{f:rho}
\end{center}
\end{figure*}

We now focus on the correlations between the two observables. Fig.~\ref{f:rho} shows the centrality dependent Pearson correlation coefficients $\rho(v_{2}^2, [p_{\rm T}])$ and $\rho(v_{3}^2, [p_{\rm T}])$, calculated by the \iebe\ model  with different initial state $\alpha$ configurations. The $\rho(v_{2}^{2}, [p_{\rm T}])$ decreases and changes from positive to negative with respect to centrality, and it decreases faster in the configurations with compact $\alpha$ cluster. The $\rho(v_{3}^{2}, [p_{\rm T}])$has negative correlations in all centralities with different nuclear densities, and it gains a strong suppression from the compact $\alpha$ cluster in oxygen. We note that the contributions of the $\alpha$ cluster to $\rho(v_{2}^{2}, [p_{\rm T}])$ have significant centrality dependence, while their contributions to$\rho(v_{3}^{2}, [p_{\rm T}])$ are weakly dependent on centrality. We therefore propose that the Pearson correlation coefficients $\rho(v_{2}^{2}, [p_{\rm T}])$ and $\rho(v_{3}^{2}, [p_{\rm T}])$ are sensitive observables to probe the compactness of $\alpha$ cluster in oxygen with relativistic heavy ion collisions.

{\em Conclusion.}
Using the \iebe\ hybrid model, we have studied the effect of $\alpha$ clusters in $^{16}$O
on the soft hadron observables in \oo\ collisions at $\snn=6.5$ TeV. 
We found that the importance of the $\alpha$ cluster for the observables in \oo\ collisions depends on the compactness of the $\alpha$ cluster (i.e. $r_{\alpha}/l$) in the light nuclei:  densities with compact cluster (small $r_\alpha/l$) give very different predictions from those of the Woods-Saxon density. The intensive skewness of the $[p_{\rm T}]$ correlator $\Gamma_{p_{\rm T}}$, the elliptic flow $v_2\{2\}$, $v_2\{4\}$, and their ratios, the triangular flow $v_3\{2\}$, the Pearson correlation coefficients $\rho(v_{2}^{2}, [p_{\rm T}])$ and $\rho(v_{3}^{2}, [p_{\rm T}])$ are sensitive to $r_\alpha/l$.
The $\alpha$ cluster effect depends on the compactness of the $\alpha$ cluster in the $^{16}$O, providing a possible way to explain the differences in previous predictions with the EQMD density and the NLEFT density. 
The magnitude of $r_\alpha/l$ reflects the properties of the strong interaction in a nucleus, which can give us some detailed information about QCD. We note that for a quantitative exploration of the compactness of $\alpha$ clusters in $^{16}$O with heavy ion collisions, more effects such as the detailed distributions of each $\alpha$ cluster, the subnucleon structure, need to be further investigated. Therefore, our study serves as an important step towards a quantitative exploration of the compactness of the $\alpha$ cluster in light nuclei in relativistic heavy ion collisions. We expect that the value $r_\alpha/l$ can be extracted from our proposed observables in the current and upcoming relativistic \oo\ collision program at RHIC and the LHC.\\

{\em Acknowledgements.}
We thanks S. Huang J. Jia, B. Lu and X. Wang  for useful discussions. This work is supported in part by the National Natural Science Foundation of China under Grant
Nos. 12247107, 12075007, HJX is supported by the National Natural Science Foundation of China under Grant Nos. 12275082, 12035006, 12075085.

\bibliography{oo}

\section{Appendix}
\begin{figure*}[!hbt]
\begin{center}
\includegraphics[width=0.95\textwidth]{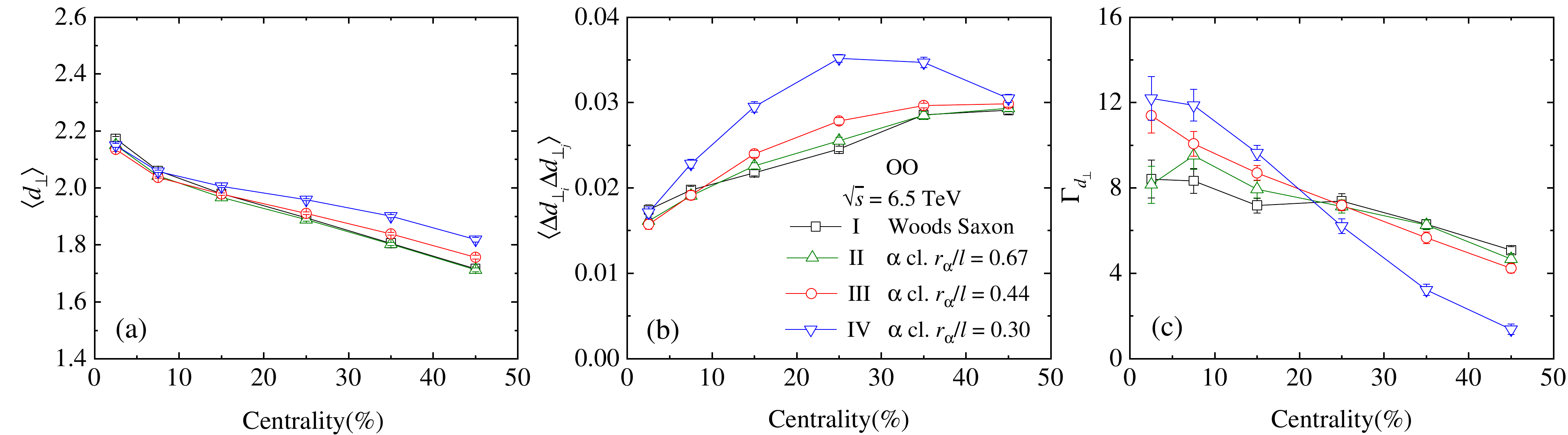}
	\caption{(Color online) The centrality dependent (a) $\langle d_{\perp}\rangle$, (b) $\langle \Delta d_{{\perp}_i} \Delta d_{{\perp}_j}\rangle $ and (c) intensive skewness of $d_{\perp}$ correlator $\Gamma_{d_{\perp}}$ in \oo\ collisions at $\snn=6.5$ TeV, calculated by the \trento\ model with different initial state $\alpha$ cluster configurations.}
\label{f:pti}
\end{center}
\end{figure*}

\begin{figure*}[!hbt]
\begin{center}
\includegraphics[width=0.95\textwidth]{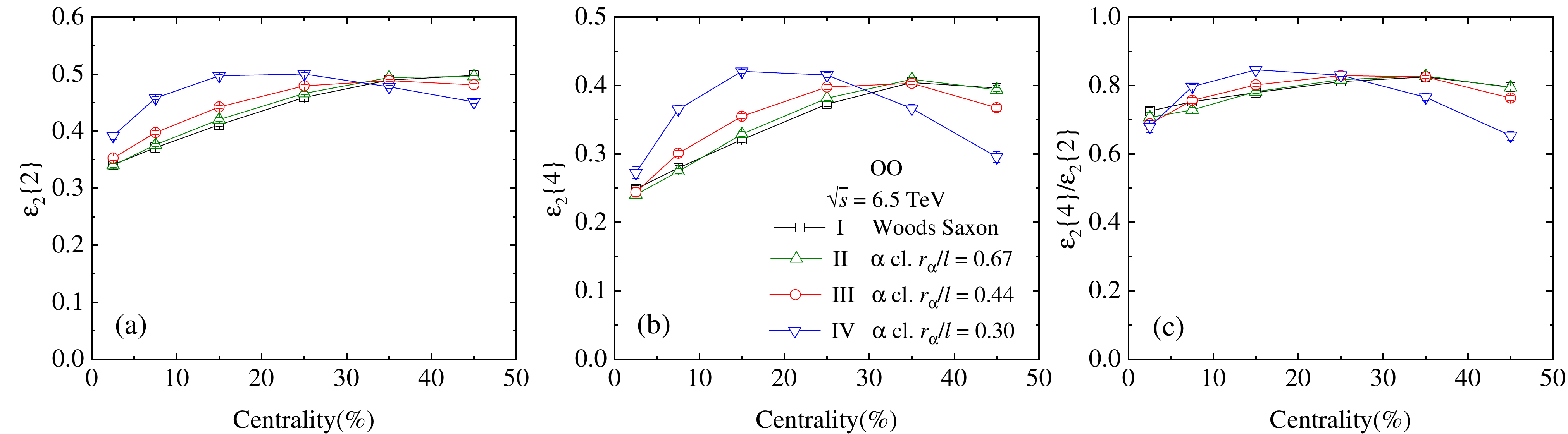}
	\caption{(Color online) The centrality dependent (a) $\epsilon_2\{2\}$, (b) $\epsilon_2\{4\}$, and (c) their ratios in $^{16}$O+$^{16}$O collisions at $\sqrt{s_{\rm{NN}}}=6.5$ TeV, calculated by  the \trento\ model.}
\label{f:v2i}
\end{center}
\end{figure*}

\begin{figure}[!hbt]
\begin{center}
\includegraphics[width=0.42\textwidth]{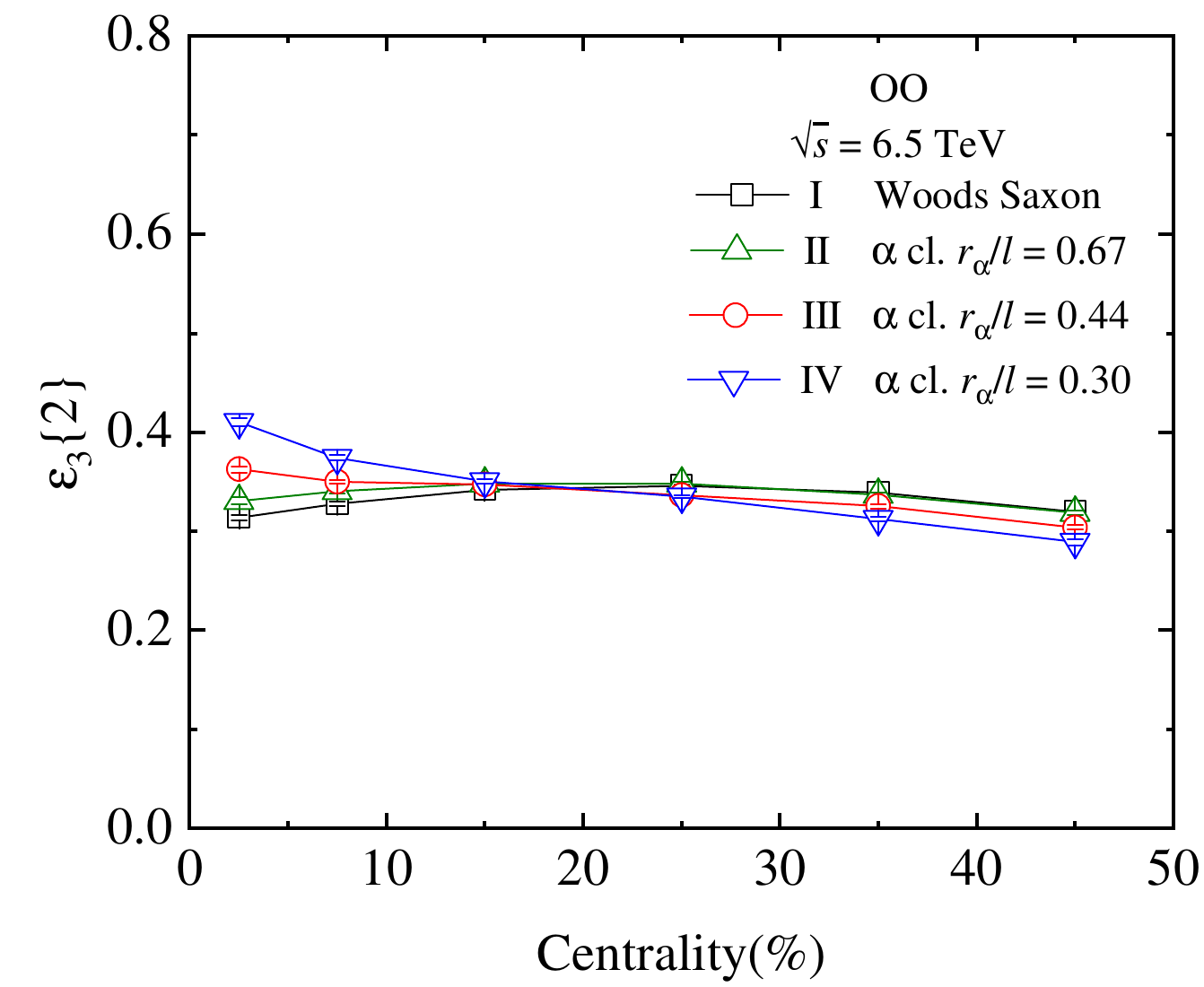}
\caption{(Color online) The centrality dependent $\epsilon_3\{2\}$ in $^{16}$O+$^{16}$O collisions at $\sqrt{s_{\rm{NN}}}=6.5$ TeV, calculated by the \trento\ model.}
\label{f:v32i}
\end{center}
\end{figure}

\begin{figure*}[!hbt]
\begin{center}
\includegraphics[width=0.9\textwidth]{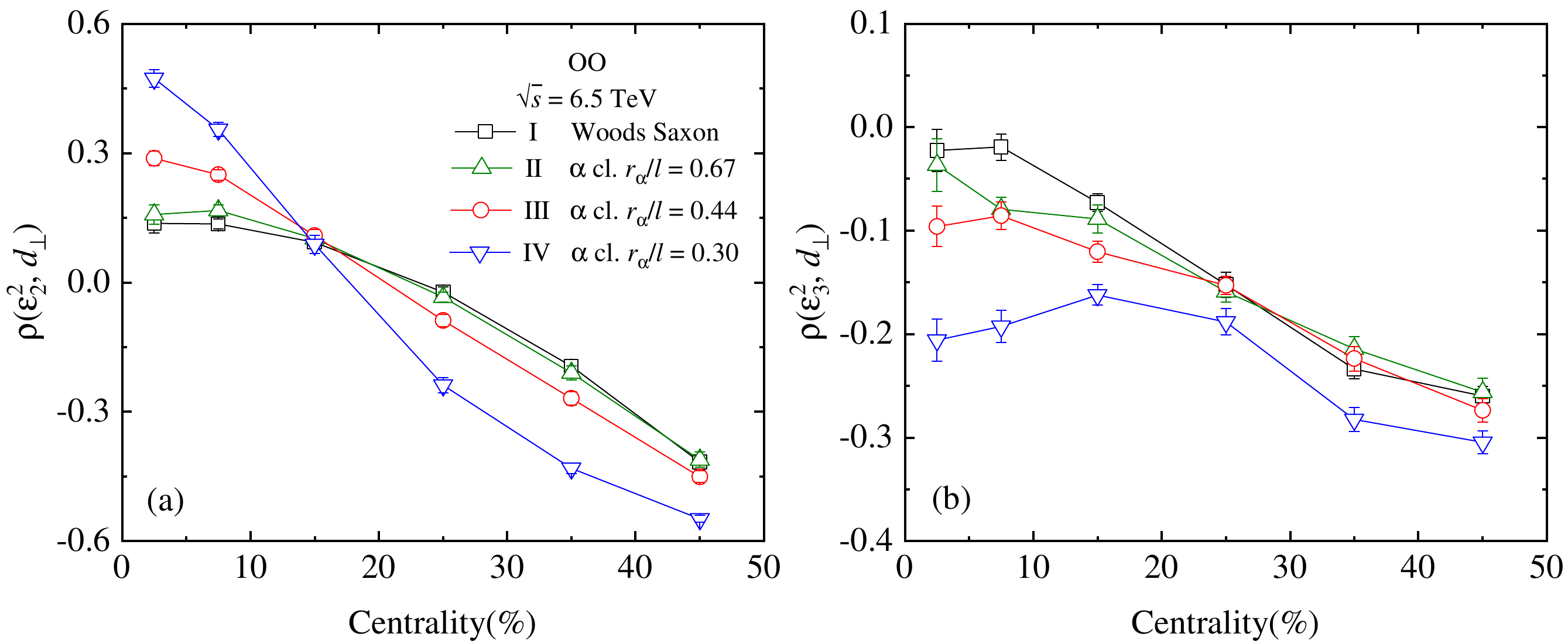}
	\caption{(Color online) The Pearson correlation coefficients (a) $\rho(\epsilon_{2}^{2}, d_{\perp})$ and (b) $\rho(\epsilon_{3}^{2}, d_{\perp})$ as a function of centrality in $^{16}$O+$^{16}$O collisions at $\sqrt{s_{\rm{NN}}}=6.5$ TeV, calculated by the \trento\ model.}
\label{f:rhoi}
\end{center}
\end{figure*}

The observables discussed in this work can be roughly described by the related initial predictors, i.e., $[p_{\rm T}]\propto d_{\perp}$ and $v_{n}\propto \epsilon_{n}$.
Here the eccentricity $\epsilon_{n}e^{in\varphi}=-\{r^{n}e^{in\varphi}\}/\{r^{n}\}$ and the initial energy per particle $d_{\perp}\equiv E/S$ are obtained from the initial profiles, where $E$/$S$ is the total initial energy/entropy. 
The initial predictors calculated from the initial \trento\ simulations are presented in Fig.~\ref{f:pti}, Fig.~\ref{f:v2i}, Fig.~\ref{f:v32i}, and Fig.~\ref{f:rhoi}, corresponding to the observables shown in Fig.~\ref{f:pt}, Fig.~\ref{f:v2}, Fig.~\ref{f:v32}, and Fig~\ref{f:rho}. 
The centralities are determined by $S$. 
All the predictors work well except for $v_{3}\{2\}$ and the Pearson coefficients $\rho(v_{n}^2,[p_{\rm T}])$, whose centrality dependence changes from initial-state predictors to final-state observables. 

\end{document}